\documentclass[twocolumn,showpacs,preprintnumbers,amsmath,amssymb]{revtex4}
\usepackage{graphicx}
\usepackage{dcolumn}
\usepackage{bm}

\begin{document}


\title{Possibility of superconductivity in 
the repulsive Hubbard model on the Shastry-Sutherland lattice} 

\author{Takashi Kimura${}^{1,2}$}
\email{kimura@cms.phys.s.u-tokyo.ac.jp}
\author{Kazuhiko Kuroki${}^3$}
\author{Ryotaro Arita${}^1$}
\author{Hideo Aoki${}^1$}
\affiliation{
${}^1$Department of Physics, University of Tokyo, 
Hongo, Tokyo 113-0033, Japan \\ 
${}^2$Department of Complexity Science and Engineering, 
University of Tokyo, Hongo, Tokyo 113-0033, Japan \\ 
${}^3$Department of Applied Physics and Chemistry, The University of
Electro-Communications, Chofu, Tokyo 182-8585, Japan 
}
\date{\today}

\begin{abstract}
Possibility of superconductivity from electron repulsion 
in the Shastry-Sutherland lattice, which 
has a spin gap at half filling, 
is explored with the repulsive Hubbard model 
in the fluctuation-exchange approximation. 
We find that, while superconductivity is not favored around the half-filling, 
superconductivity is favored around the quarter-filling. 
Our results suggest that the Fermi surface nesting is 
more important than the spin dimerization for superconductivity. 
\end{abstract} 

\maketitle

Superconductivity from electron repulsion 
has acquired a renewed momentum  
from the discovery of high-$T_C$ cuprates \cite{Bednorz}, 
and the subsequent seminal proposal by Anderson \cite{Anderson} that 
the electron correlation should be at the heart of the 
superconductivity. 
An important, and still not fully understood, question 
dating back to the early stage along this line 
is the relation between the spin gap 
and the superconductivity.  
A gap in the spin excitation associated with a quantum spin liquid 
is, crudely speaking, favorable for a singlet-pairing formation. 
So one may naively expect that doping of carriers 
into a spin-gapped Mott insulator will show superconductivity. 
A clear-cut example is the $t$-$J$ or Hubbard model 
on the two-leg ladder\cite{ladder}, 
where the un-doped ladder is a spin-gapped Mott insulator 
and the system indeed becomes superconducting when doped.  

On the other hand, the spin gap is obviously not a {\it necessary} 
condition for superconductivity. 
The Hubbard model on the 2D square lattice 
or the three-leg ladder \cite{three-leg} 
are examples, which are spin-gapless 
at half-filling\cite{Heisenberg,ladder}, 
but superconduct when doped\cite{three-leg}. 

If we go back to the superconductivity from 
repulsive electron-electron interactions in a broader context, 
usual understanding is that the effective attraction 
between electrons are mediated by spin fluctuations.\cite{Scalapino,SCR} 
An important difference from the superconductivity from 
attraction is that the effective attraction arises from 
pair-scattering processes across which the BCS gap 
function changes sign, so the effective attraction 
is wave-number dependent and 
the pairing is anisotropic (typically $d$-wave as in the cuprates). 
Numerically, a quantum Monte Carlo calculation that takes 
care of the relevant energy scale\cite{Kuroki&Aoki,Aokireview} 
has shown an enhanced pairing correlation in the repulsive Hubbard model. 
Analytically, the fluctuation-exchange (FLEX) studies 
\cite{Bickers,Grabowski,Dahm}, 
which is a kind of the renormalized random phase approximation 
based on the Fermi liquid picture, 
have shown $d$-wave superconductivity with 
a transition temperature $T_C\sim O(0.01t)$ 
for the repulsive Hubbard model.  
Notably, $T_C$ is two orders of magnitude smaller 
than $t$, although $T_C$ amounts to $\sim 100$K as 
in the high-$T_C$ cuprates if we take $t\sim 0.4$ eV.  

Recently, the multiband lattices having 
{\it disconnected Fermi surfaces} 
have been proposed as systems having much higher $T_C$'s 
($\gg 0.01t$) \cite{KurokiArita,NTT,KTA,Zeni,Onari}.  
While the motivation 
for considering disconnected Fermi surfaces is to raise 
the $T_C$, which is lowered down to $\sim 0.01t$ 
in ordinary lattices because the nodes in the BCS gap function 
intersect the Fermi surface: when the Fermi surface consists 
of pockets, the nodes can run in between the pockets.  
If we look at the proposed lattices in real space, on 
the other hand, we immediately notice 
that all of them happen to have dimerized structures in 
some way or other.  
From the spin-gap point of view one might consider this 
reasonable, since the dimerization 
can favor a spin gap when the dimerization is strong. 
However, we cannot identify 
which of the dimerization and the disconnected 
Fermi surface contributes to the higher $T_C$, 
since the dimerization and the disconnected 
Fermi surface are simultaneously satisfied in these lattices. 

Now, there is an important and unresolved question of 
whether the spin gap associated with strong dimerization 
is a {\it sufficient} condition for superconductivity, 
i.e., whether the doped spin-gapped system can 
always become superconducting with an appreciable $T_C$. 
There are several lattices intensively studied from the viewpoint 
of the spin gap in the Mott-insulator phase. 
Among them is the Shastry-Sutherland (SS) lattice (Fig. 1), 
where we have a herringbone array of dimers.  
This lattice is first proposed by Shastry and Sutherland \cite{SS} 
about two decades ago. They found that 
the ground state in the Heisenberg model is spin-gapped 
when strongly dimerized, $J_2/J_1<0.5$, 
where $J_1(J_2)$ is the 
spin-spin interaction within (across) dimer. 
With the exact diagonalization method for finite clusters,  
Miyahara and Ueda \cite{MiyaharaUeda} have recently 
obtained a more accurate critical value, $J_2/J_1 \simeq 0.7$, 
for the appearance of the spin gap.  Analytic studies such as 
a series expansion \cite{Koga}, rigorous bounds \cite{Low}, 
and large-$N$ theories \cite{Chung} have indicated similar boundaries. 
\begin{figure}
\begin{center}
\includegraphics[width=0.5\linewidth]{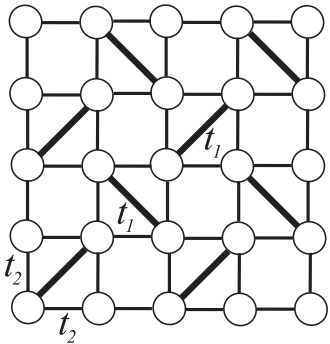}
\caption{
The Shastry-Sutherland lattice. 
}
\label{fig1}
\end{center}
\end{figure}

A recent impetus came from experiments on a copper compound, 
SrCu$_2$(BO$_3$)$_2$,\cite{Kageyama1} where the Shastry-Sutherland lattice 
is realized. 
Experimental results for magnetic susceptibility \cite{Kageyama1,Kageyama2}, 
Cu NQR \cite{Kageyama1}, 
high-field magnetization \cite{Kageyama1}, 
ESR \cite{Nojiri}, Raman scattering \cite{Lemments}, 
inelastic neutron scattering \cite{Kageyama3} and specific heat 
\cite{Kageyama4} 
show that this compound 
has a dimerized ground state with 
a spin gap $\simeq 30$K with no long-range magnetic order.\cite{plateau} 

So the SS lattice is an appropriate system for the 
above question of whether a spin-gapped system can superconduct 
when doped.  
A recent study \cite{ShastryKumar} for the $t$-$J$ model on the SS lattice 
has indeed discussed possible superconducting transition 
at low temperatures of $O(0.01t)$ 
based on RVB type mean field theory \cite{RVB}. 
Usually the SS lattice is studied with 
the Heisenberg model, which corresponds to the half-filled 
Hubbard model.  Here we have opted for the Hubbard model, 
since we can only study a finite Hubbard $U$, but 
also look at the SS lattice around quarter filling, 
which we propose here to be interesting.  
The band filling controls the shape of the Fermi surface after all, 
so that it should be an important parameter for studying 
the question at hand.
Quarter-filled case is not unrealistic, since the herringbone structure of 
the SS lattice strongly reminds us 
of a class of dimerized organic crystals, 
where the band is often {\it quarter filled} 
rather than half-filled. 

So here we take the SS lattice to study superconductivity in the Hubbard 
model.  We adopt the FLEX approximation, which has 
to be extended to four-band systems\cite{NTT,KTA,Zeni,KurokiBEDT} 
to treat the SS lattice that has four atoms per unit cell. 
Superconducting transition has been examined 
with Eliashberg's equation \cite{Eliashberg}. 
We shall show that the Hubbard model, around 
half filling, does not exhibit 
superconductivity with significant $T_C$.
So this provides an example in which a doped spin-gapped system 
does not guarantee an appreciable $T_C$. 
By contrast, the Hubbard model around the quarter-filling 
exhibits superconductivity with a $d$-wave pairing 
when the dimerization is strong $t_2\ll t_1$.  
We shall discuss this in terms of the shape of the Fermi surface 
and the pair-scattering processes on it.  

In the four-band version of the FLEX \cite{NTT,KTA,Zeni,KurokiBEDT}, 
Green's function $G$, spin susceptibility $\chi$, 
self-energy $\Sigma$ are $4\times4$ matrices, e.g., 
$G_{lm}({\vec k}, i\epsilon_p)$ 
with $\epsilon_p$ being the Matsubara frequency.  
Here $l,m$ refer to the four sites in a unit cell, which 
can be unitary-transformed to band indices. 
We obtain the eigenvalue and the superconducting gap function $\phi_{lm}$ 
by solving the linearized Eliashberg's equation \cite{Eliashberg}.  
For the susceptibility $\chi$, we quote hereafter the value of the 
largest component when we diagonalize its matrix.  
In the present study, we take $32\times32$ $k-$points and  
up to 8196 Matsubara frequencies, or $64\times64$ $k-$points and  
up to 4096 Matsubara frequencies. 

We start with the case of near half-filling. 
Here we take the band filling 
$n$(=number of electrons/number of sites)=0.85, along with the 
on-site repulsion $U=7$ and the 
transfer energy within (across) dimer $t_1=\pm 1.25$ ($t_2=1.0$). 
This value of $t_1$ has been adopted in
Ref.\cite{ShastryKumar} as appropriate to SrCu$_2$(BO$_3$)$_2$.  
Corresponding Heisenberg model at half filling 
has a spin gap because of $J_2/J_1\sim t_2^2/t_1^2 \ll 0.7$. 

The maximum eigenvalue of Eliashberg's equation $\lambda$ 
has turned out to be much smaller than unity 
($\lambda\simeq 0.43$) at low temperature ($0.01\le T\le0.04$).  
We have also calculated for $t_1=1.8$, which is a 
favorable value for the near-quarter-filled case 
as we shall see below, but $\lambda$ is again very small 
($\lambda=$0.36, 0.37 at $T=$0.04, 0.01, respectively). 

\begin{figure}
\begin{center}
\includegraphics[width=\linewidth]{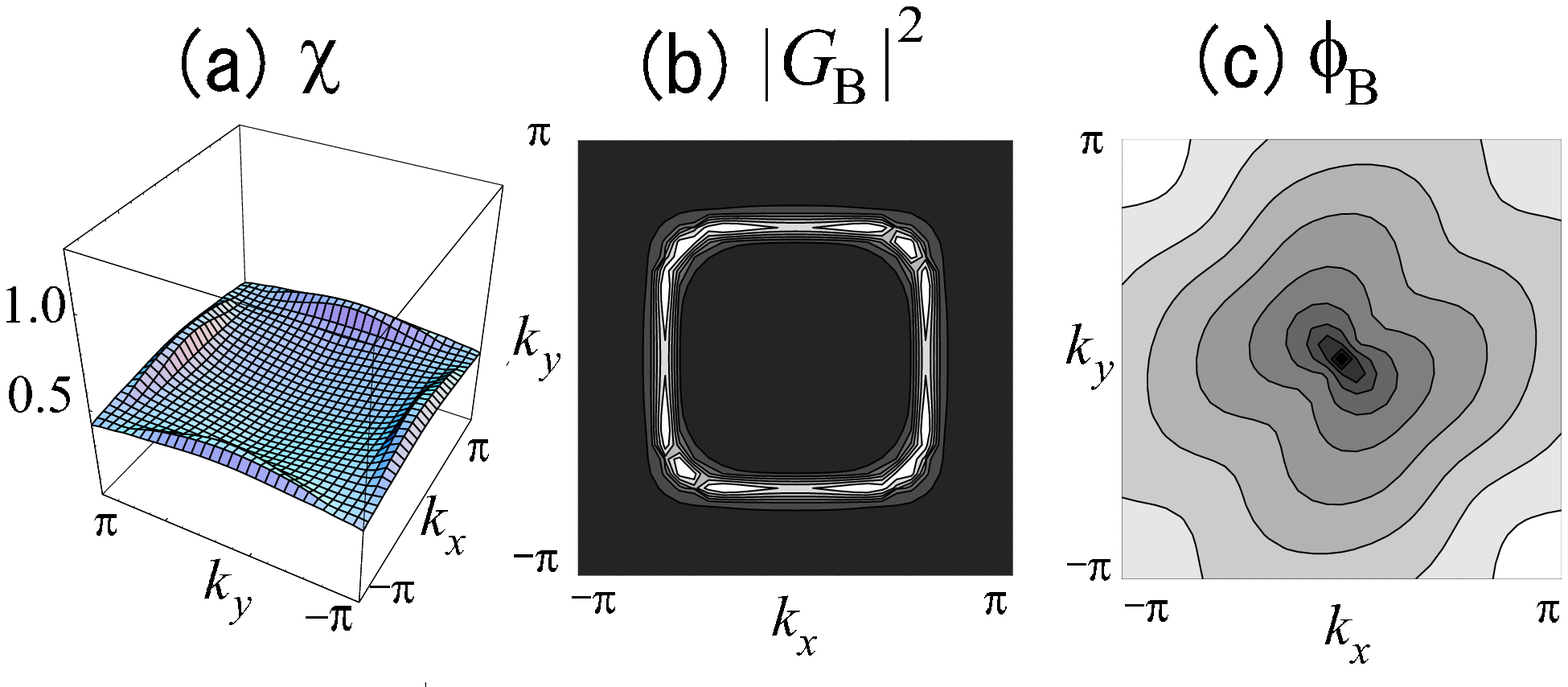}
\caption{
Result for $\chi$ (a), 
$|G_B|^2$ (b), and $\phi_B$ (c) against $k_x, k_y$ for the 
lowest Matsubara frequency 
for the nearly half-filled $n=0.85$ with 
$U=7$, $t_1=1.8$, $t_2=1.0$, and $T=0.025$. 
}
\label{fig2}
\end{center}
\end{figure}
\begin{figure}
\begin{center}
\includegraphics[width=1.0\linewidth]{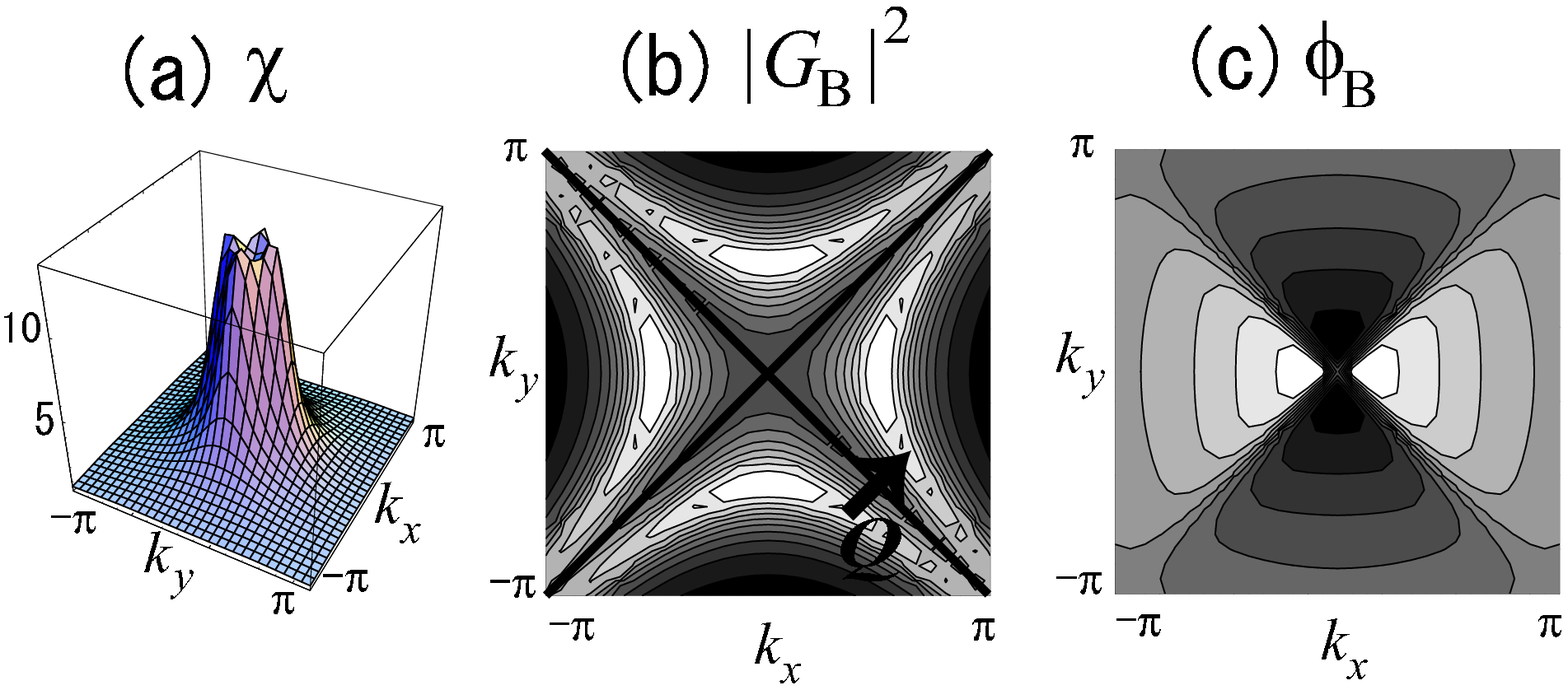}
\caption{
The same plot as in Fig. 2 for a weaker dimerization with $t_1=0.5$. 
${\bf Q}$ represent the nesting vector across
$k_x=\pm k_y$ (solid lines).
} 
\label{fig3} 
\end{center} 
\end{figure} 
Figure 2 shows the spin susceptibility 
($\chi$), Green's function ($G$) and the gap function ($\phi$)
for the second band from the bottom 
(called $B$, which crosses the Fermi energy) for 
$t_1=1.8$ and $t_2=1.0$ at $T=0.025$. 
We see the susceptibility has no strong peaks, 
unlike in the square lattice which has a large 
antiferromagnetic peak in the susceptibility around $\Vec{k}=(\pi,\pi)$ 
that is relevant to the superconductivity.  
The weak spin structure in the SS lattice 
should be due to the spin gap and/or 
the spin frustration which strongly prevent(s)
the long range spin correlation when the dimerization is strong\cite{Note}.   

If we weaken the dimerization by making $t_1$ 
sufficiently smaller than $t_2$, the system becomes superconductive.
Figure 3 shows the spin susceptibility, 
Green's function and the gap function 
for $t_1=0.5$ and $t_2=1.0$ at $T=0.025$, where we obtain 
a large $\lambda=0.94$ close to unity ($\lambda$ becomes unity  
at lower temperature $T\sim 0.02$). 
In the figure 
we see the Fermi-surface nesting across $k_x=\pm k_y$ is 
appreciable and the spin susceptibility has a strong peak around $(0,0)$, 
which corresponds to the peak around $(\pi,\pi)$ 
on the square lattice ($t_1=0$) folded. 
The result indicates that the dimerization (or the spin gap) 
is by no means a sufficient condition for 
very high $T_C$. 
If we turn to the very high $T_C$ systems obtained in \cite{KurokiArita,NTT,KTA,Zeni,Onari}, dimerization causes the disconnected 
Fermi surfaces accompanied by 
antiferromagnetic spin fluctuations. 
So the present result is an example in which 
a dimerization works unfavorably for superconductivity 
in a simply-connected Fermi surface.  
In this sense, the disconnected Fermi surfaces rather than the dimerization 
is essential for very high $T_C$ in \cite{KurokiArita,NTT,KTA,Zeni,Onari}.  

We have a drastically different situation when we change 
the band filling to quarter filling.  
Figure 4 shows the temperature dependence of 
$\lambda$ at $n=0.55$ with the same parameter as those at 
$n=0.85$. We see $\lambda$ is strongly enhanced at low temperatures.

Figure 5 shows $\chi$, $G$ and $\phi$ for the lowest two bands 
(called A and B) that cross the Fermi energy for this filling. 
The peak in the spin susceptibility around $(0,0)$
is much stronger than for the half-filling.  The spin fluctuation 
should mediate the pair-scattering across 
which the gap function has opposite signs, 
resulting in $d$-wave superconductivity.  
How can this happen when the nesting vector is 
close to $(0,0)$?  Figure 5(d) depicts the answer: 
the gap function has a $d$-wave symmetry
as in the square lattice near half-filling. 
This is physically natural because 
the SS lattice around quarter filling is 
effectively a square lattice around half filling 
in the strongly dimerized case $|t_1|\gg|t_2|$.  
\begin{figure}
\begin{center}
\includegraphics[width=0.7\linewidth]{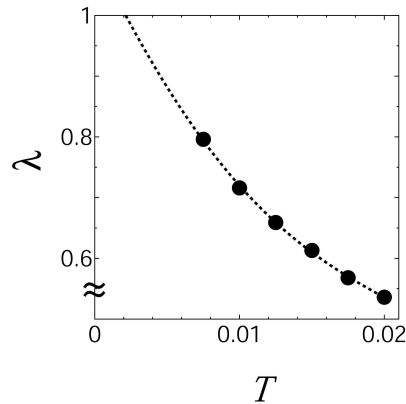}
\caption{ 
For a nearly quarter-filled band ($n=0.55$) 
$\lambda$ is plotted as a function of temperature $T$ 
for $U=7$, $t_1=1.8$ and $t_2=1.0$.  
The dotted curve is a least-square 
fit by a fourth-order polynomial. 
Here we take $32\times 32$ $k$-points and up to 
8196 Matsubara frequencies.  
Error bars for different choice of the 
$k$-point mesh and Matsubara frequencies 
are $\sim 10\%$ at low temperatures. 
}
\label{fig4}
\end{center}
\end{figure}
\begin{figure}
\begin{center}
\includegraphics[width=0.7\linewidth]{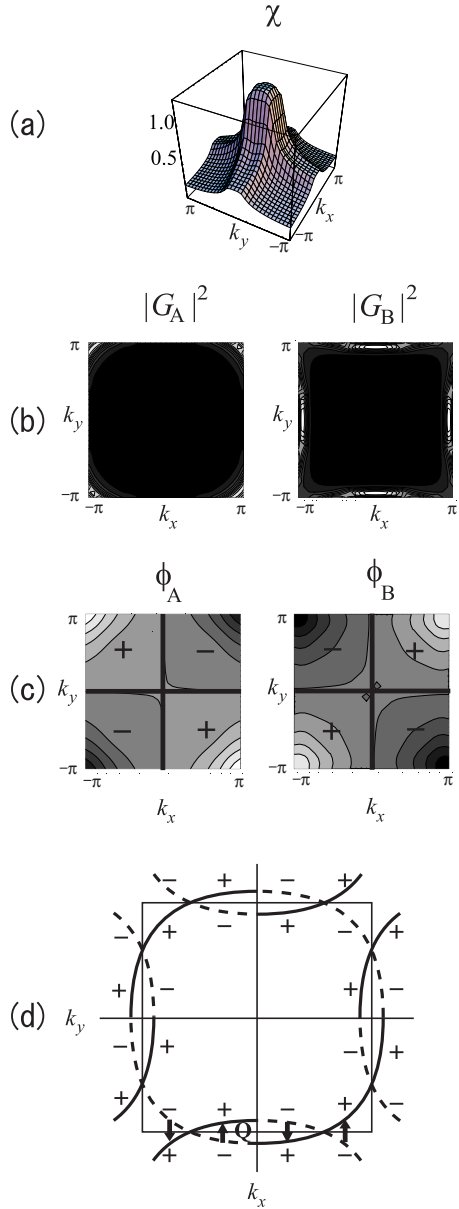}
\caption{ 
For a nearly quarter-filled band ($n=0.55$) 
$\chi$ (a), $|G_\nu|^2$ (b), 
and $\phi_\nu$ (c) ($\nu=A,B$)
are plotted against $k_x, k_y$ for the 
static case (lowest Matsubara frequency) 
for $U=7$, $t_1=1.8$, $t_2=1.0$, and $T=0.025$. 
(d) schematically depicts the sign of the gap function 
on the Fermi surface, where ${\bf Q}$ and arrows 
represent the nesting vector $\sim (0,0)$. 
}
\label{fig5}
\end{center}
\end{figure}

As mentioned above, it is interesting to compare the SS lattice 
with organic, $d$-wave superconductors such as 
$\kappa$-(BEDT-TTF)$_2$X.\cite{BEDT}  
If we assume the dimerization is sufficiently strong, 
the original system around quarter filling  
can be represented by a two-band\cite{Schmalian,KurokiAokiBEDT} 
or a single-band\cite{KinoKontani,onebandBEDT} 
Hubbard model around half filling.\cite{KurokiBEDT}  
There, superconductivity is enhanced 
for $\sim 1/4$ filled band 
by strong dimerization as in the present study. 
Although organic materials that can be 
modeled by the SS lattice have not been known, 
it would be interesting to search for them. 

In summary, we have studied superconductivity in 
the Hubbard model on the Shastry-Sutherland lattice with FLEX. 
Our analysis shows superconducting transition temperature, 
if any, is very small around half filling despite the 
presence of a spin gap due to the dimerization, 
while superconductivity is favored around quarter filling. 
Comparison with the RVB theory
for the $t$-$J$ model\cite{ShastryKumar}
is interesting because the RVB theory has shown $T_C$ of 
$O(0.01t)$.  While this may naively seem inconsistent with our result, 
the $T_C$ of a conventional RVB theory 
on the SS lattice is much smaller than that ($\sim O(0.1t)$) for 
the square lattice\cite{tJ,Ubbens}.  
So the situation about the difference between $T_C$'s 
is similar to this case.  

This work was supported in part by a Grant-in-Aid for Scientific 
Research and Special Coordination Funds from the Ministry of 
Education of Japan. 
Numerical calculations were performed at the Supercomputer Center,
Institute for Solid State Physics, University of Tokyo.



\begin{thebibliography}{99}
\bibitem{KurokiBEDT} 
To be precise, the pairing symmetry 
should be determined by the original four-band model as shown by 
K. Kuroki, T. Kimura, R. Arita, Y. Tanaka, and Y. Matsuda 
[Phys. Rev. B {\bf 65}, 100516 (2002)].
\bibitem{Note} Generally, it is hard to identify which condition 
is essential for preventing the long-range spin correlation. 
For example, the spin frustration prevents the spin correlation 
in many systems 
but the weak next-nearest hopping in the square lattice 
does not strongly prevents the spin correlation 
and even enhances the superconductivity \cite{KinoKontani}. 
On the other hand, the spin gap also 
prevents the spin correlation in many systems
such as ladder systems, but does not result in very high $T_C$ systems.
\end{thebibliography}
\end{document}